\documentclass[12pt]{article}
\usepackage{amsfonts}
\usepackage[latin1]{inputenc}
\usepackage[nooneline,up]{subfigure}
\usepackage{graphicx}
\title{Generalized Heisenberg Algebras and Fibonacci Series}
\author{J. de Souza, E. M. F. Curado and M. A. Rego-Monteiro\\
Centro Brasileiro de Pesquisa F\'{\i}sicas\\
Rua Dr. Xavier Sigaud, 150\\
22290-180 - Rio de Janeiro, Brazil}
\date{}
\begin{document}
\maketitle

\begin{abstract}

\indent
We have constructed a Heisenberg-type algebra generated by the 
Hamiltonian, the step operators and an auxiliar operator. This algebra 
describes quantum systems having eigenvalues of the
Hamiltonian depending on the eigenvalues of the two previous 
levels. This happens, for example, for systems having the energy spectrum 
given by Fibonacci sequence. Moreover, the algebraic structure depends on two 
functions $f(x)$ and $g(x)$.  When these two functions are linear we 
classify, analysing the stability of the fixed points of the functions, the possible representations for this algebra.  

\end{abstract}

\vspace{0.5cm}

\begin{tabbing}

\=xxxxxxxxxxxxxxxxxx\= \kill

Fock representation.
\\


\end{tabbing}

\section{Introduction}

Due to their interesting mathematical structure and possible physical applications
deformed algebras have attracted much attention in the last twenty years. In 1982 
Kulish \cite{kr} showed that 
the underlying algebra of the $XXZ$-Heisenberg spin model was a deformation of the 
$su(2)$ algebra, called nowadays $su_{q}(2)$. 

Following the technique developed by Schwinger to compose two Heisenberg 
oscillators to obtain $su(2)$ algebra, 
in 1989 Macfarlane and  Biedenharn \cite{mf,b} constructed the 
$su_{q}(2)$ algebra using two $q$-oscillators. $q$-oscillator algebra is a deformation 
of the Heisenberg algebra through a parameter $q$, which reproduces Heisenberg algebra when the parameter $q \rightarrow 1$.

Since Heisenberg algebra has an important role in several areas of physics
it was tempting to search for applications of this new structure. Thus, there was an 
intense study of possible physical applications of $q$-oscillators and since then some 
succesfull ones have been described in the literature \cite{mi,adf,angelova0,bd,mrw,lavagno,
algin,nmon}. 
They have also been used  as a phenomenological approach to study 
composite particles  
\cite{jnc,bon1,bcr1,bcr2,bcr3}, spectra of atoms and molecules 
\cite{jnc,bon-rot,bon-vib,angelova0,angelova2,nmon}, generalized field theory 
\cite{fin,bcr1,bcr2,bcr3} and coherent states \cite{hcr}. 

Recently, a new algebraic structure was proposed \cite{cr} that generalizes the 
Heisenberg algebra. This structure contains also $q$-oscillators as a particular case 
and has been successfully used in some different physical problems \cite{nmon,
jnc,bcr1,bcr2,bcr3,bcr4,hcr,hcrl}.
In this new algebra, called Generalized Heisenberg Algebra (GHA), 
the commutation relations among the operators $a$, $a^{\dagger}$ and $J_0$ depend on 
a general analytic functional $f(x)$ and  are given by:
\begin{equation}\label{eq0a}
J_0a^\dagger =a^\dagger f(J_0)
\end{equation}
\begin{equation}\label{eq0b}
a J_0=f(J_0)a 
\end{equation}
\begin{equation}\label{eq0c}
[a ,a^\dagger ]=f(J_0)-J_0,
\end{equation}
with  $a =(a^{\dagger})^{\dagger}$ and $J_0=J_{0}^{\dag}$. 
Identifying the operator $J_0$ with the Hamiltonian of a
physical system, this algebra tell us that  
the $J_0$ eigenvalues ($J_0|n\rangle=e_n|n\rangle$ in Fock space), which 
are the energy spectrum of the physical system, are obtained by an 
one-step recurrence ($e_n=f(e_{n-1})$), {\it i.e}, each eigenvalue depends on the 
previous one.  Thus, the eigenvalue behavior can be studied by dynamical system 
techniques, enormously simplifying the task of finding possible representations of the 
algebra \cite{cr}. 

When the functional $f(x)$ 
is linear we re-obtain the $q$-oscillator algebra \cite{cr,bl}; other kinds of functionals 
give more general 
algebraic structures than $q$-oscillators \cite{cr}. The functional $f(x)$ can be, for 
example, a polynomial and therefore it depends on some parameters (the polynomial coefficients). 
Depending on the kind of functional and their parameters, finite or infinite representations 
are allowed \cite{cr}. The $J_0$ operator eigenvalues depend on the characteristic function 
$f(x)$ and it can  present a variety of different behaviors (monotonously increasing or decreasing,  
etc, upper limited or not, etc) \cite{cr}. For instance, for the functional $f(J_0)=\left(\sqrt{J_0}+\sqrt{b}\right)^2$, where $b=\pi^2/2mL^2$
with  $m$ and  $L$ being the mass and the length of the well, one obtains the square-well potential 
algebra \cite{crn}. However, there are several systems which cannot be described by 
such a class of algebras. 

Quasi-periodic structures have attracted much interest lately
\cite{albuquerque}. In particular, Fibonacci sequences were 
considered in several areas of mathematics and physics 
\cite{albuquerque,paperjef1}.
Moreover, it was shown that an electron gas under sudden
heating has a quasiequilibrium regime with a
quasiperiodic energy distribution \cite{prl}.  Since most systems 
having a quasiperiodic energy spectrum cannot be described by the GHA,
in this paper we are going to present a first step in this direction. 
We construct here a Heisenberg-type algebra for a system having
the energy spectrum described by a generalized Fibonacci
sequence. In order to realize this, we  propose a simple
algebraic structure depending on two functionals, $f(x)$ and 
$g(x)$. We introduce an operator $H$, being the Hamiltonian of 
a system \footnote{Instead of calling $J_0$ as it was done in Eqs. 
(\ref{eq0a} - \ref{eq0c}), we will call this operator, from now on, directly 
as $H$.}, 
having eigenvalues obeying 
the equation $e_{n+1}=f(e_{n},e_{n-1})$, 
i.e., the eigenvalue of the $n$-th state depends on the two previous 
eigenvalues. We call this structure a {\it two-step algebra}. In order to extend 
the ``one-step algebra'' described by the GHA, we introduce also
an additional operator $J_3$.

When both functionals $f(x)$ and $g(x)$ appearing in the two-step algebra are linear functions, we show that we can re-write both the $H$ and  $J_3$ operators in Fock space 
as functions of a single  number operator, $N$. The algebraic structure generated by these functionals contains the GHA, $q$-oscillator algebra and is related to 
other interesting algebras as special cases. 
 
 In section II we introduce the two-step generalized Heisenberg algebra. In section III 
 we study the linear case for the algebra, that contains a Fibonacci-like spectra and 
 many other interesting quasi-periodic sequences and we also discuss the types of 
 representations we could find. 
  In section IV we present our conclusions and in the appendix we classify 
 possible representations for the algebra when $f(x)$ and $g(x)$, appearing in 
 the algebra, are linear functions.

\section{Extended Two-step Heisenberg Algebra}

We propose an structure generated by the operators $H$, $J_3$, $a^{\dag}$ and $a$, 
with $a$ and $a^{\dagger}$, $a =(a^{\dag})^{\dag}$, being the step operators, 
$H=(H)^\dag$ the Hamiltonian 
 and $J_3=(J_3)^\dag$, obeying the following relations:

\parbox[t][1.cm][c]{4cm}{\[ \hspace{3.cm}\left\{ \begin{array}{cc}
Ha^\dagger &= \\
a H &= 
\end{array} \right. \] } 
\parbox[t][1.6cm][b]{6cm}{
\begin{eqnarray}
&a^\dagger (f(H)+J_3);\label{eq1} \\
&(f(H )+J_3)a ;\label{eq2} 
\end{eqnarray}
}

\parbox[t][1.cm][c]{4cm}{\[ \hspace{3.cm}\left\{ \begin{array}{cc}
J_3a^\dagger &= \\
a J_3 &=
\end{array} \right. \] } 
\parbox[t][1.6cm][b]{6cm}{
\begin{eqnarray}
&a^\dagger g(H);\label{eq3}\\
&g(H)a ;\label{eq4}
\end{eqnarray}
}

\parbox[t][1.cm][c]{4cm}{\[ \hspace{3.cm}\left\{ \begin{array}{cc}
\left[ a ,a^\dagger \right]&=\\
\left[H,J_3\right]&=
\end{array} \right. \] } 
\parbox[t][1.6cm][b]{6cm}{
\begin{eqnarray}
&f(H)-H+J_3;\label{eq5}\\
&0\label{eq6},
\end{eqnarray}
}\\
where we have assumed that $f(x)$ and $g(x)$ are analytical functions. In the Fock space representation of this structure, one has the normalized vacuum state 
$|0 \rangle$ defined by the relations:
\begin{eqnarray}
a |0  \rangle&=&0\label{eq7a}\\
H|0  \rangle&=&\alpha_0|0  \rangle\label{eq7b}\\
J_3|0  \rangle&=&\beta_0|0  \rangle\label{eq7c} ,
\end{eqnarray}
where $\alpha_0$ and $\beta_0$ are real numbers. 

A Casimir operator associated with this algebraic relations is
\begin{equation}\label{eq7.5}
C^{(1)}=a a^\dagger -f(H)-J_3=a^\dagger a -H,
\end{equation} 
satisfying the following relations: 
\begin{equation}\label{eq7.5a}
[C^{(1)},a^\dagger ]=[C^{(1)},a ]=[C^{(1)},H]=[C^{(1)},J_3]=0.
\end{equation}

\subsection{Representation Theory}

In order to build a representation theory for the structure given by (\ref{eq1}-\ref{eq6}), let us start with the vacuum state defined by relations (\ref{eq7a}-\ref{eq7c}). 

The $a^\dagger $ operator acting on the vacuum state produces another vector, say $|k  \rangle$:  $a^\dagger |0  \rangle=|k  \rangle$. It is trivial to see that the new vector $|k  \rangle$ is  orthogonal to vector $|0  \rangle$. Let us call $|k  \rangle = N_0|1  \rangle$. The $N_0$ constant can be  determined by performing the inner product  of  $|k \rangle$ and its dual and by using relation (\ref{eq5}):
\begin{equation}\label{eq11}
\langle k|k  \rangle = \langle 0|a a^\dagger |0  \rangle = \langle 0|a^\dagger a +f(H)-H+J_3|0  \rangle \Rightarrow N_{0}^2=f(\alpha_0)-\alpha_0+\beta_0.
\end{equation} 

In order to determine the $H$ eigenvalue on the state  $|1 \rangle$, we have, using equation (\ref{eq1}):
\begin{equation}\label{eq13}
H|1 \rangle \equiv \alpha_1 | 1 \rangle =  (f(\alpha_0)+\beta_0)|1 \rangle.
\end{equation}

Just as $H$, we can calculate the $J_3$ eigenvalue using equation (\ref{eq3}),
\begin{equation}\label{eq15}
J_3|1 \rangle \equiv \beta_1|1 \rangle \Rightarrow J_3\frac{a^\dagger }{N_0}|0 \rangle = \frac{a^\dagger }{N_0}g(H)|0 \rangle = g(\alpha_0)|1 \rangle,
\end{equation}
therefore, we have $\beta_1=g(\alpha_0)$.

In general, it can be proved that:
\begin{eqnarray}
H|n \rangle = &\alpha_{n}|n \rangle;\label{eq28c}\\
J_3|n \rangle = &\beta_{n}|n \rangle,\label{eq28d},\\
a^\dagger |n \rangle = &N_{n}|n+1 \rangle;\label{eq28a}\\
a |n \rangle = &N_{n-1}|n-1 \rangle;\label{eq28b}
\end{eqnarray}
where
\begin{eqnarray}
\alpha_{n+1}=&f(\alpha_{n})+\beta_{n},\label{eq29a}\\
\beta_{n+1}=&g(\alpha_{n}),\label{eq29b}\\
N_{n+1}^2=& N_{n}^{2}+f(\alpha_{n+1})-\alpha_{n+1}+\beta_{n+1}\label{eq29c}\\
=&\alpha_{n+2}-\alpha_{0}.\nonumber
\end{eqnarray}
If we combine equations (\ref{eq29a}) and (\ref{eq29b}), the two-step dependence of the $H$ eigenvalue becomes clear: $\alpha_{n+1}=f(\alpha_{n})+g(\alpha_{n-1})$, where $\alpha_0\le \alpha_{n}$, for all $n>0$. Choosing different functions and different initial values $\alpha_0$ and $\beta_0$, we can construct different Fock space representations for this mathematical structure. 
In matrix representation, the operators of the algebra can be written as
\begin{eqnarray}
    H = \left(  
    \begin{array}{ccccc}
        \alpha_{0} & 0 & 0 & 0 & \ldots  \\
        0 & \alpha_{1} & 0 & 0 & \ldots  \\
        0 & 0 & \alpha_{2} & 0 & \ldots  \\
        0 & 0 & 0 & \alpha_{3} & \ldots  \\
        \vdots & \vdots & \vdots & \vdots & \ddots 
    \end{array}
    \right)   ,&
 J_{3}= \left(  
    \begin{array}{ccccc}
        \beta_{0} & 0 & 0 & 0 & \ldots  \\
        0 & \beta_{1} & 0 & 0 & \ldots  \\
        0 & 0 & \beta_{2} & 0 & \ldots  \\
        0 & 0 & 0 & \beta_{3} & \ldots  \\
        \vdots & \vdots & \vdots & \vdots & \ddots 
    \end{array} 
    \right), \nonumber \\  
       a^{\dagger}= \left(  
    \begin{array}{ccccc}
        0 & 0 & 0 & 0 & \ldots  \\
        N_{0} & 0 & 0 & 0 & \ldots  \\
        0 & N_{1} & 0 & 0 & \ldots  \\
        0 & 0 & N_{2} & 0 & \ldots  \\
        \vdots & \vdots & \vdots & \vdots & \ddots 
    \end{array}
    \right) ,&
    a =  (a^{\dagger})^{\dagger}  .
    \label{eq:matriz}
\end{eqnarray}

In the next section, we shall study the case where $f(x)$ and $g(x)$ are linear functions.

\section{The linear case - Generalized Fibonacci Algebra}

If we assume $f(x)=rx$ and $g(x)=sx$ ($r, s \in \mathbb{R}$), the equations 
(\ref{eq1}-\ref{eq6}) reduce to
\begin{eqnarray}
Ha^\dagger =&a^\dagger (rH+J_3) \label{eq53a}\\
a H=&(rH+J_3)a \label{eq53b}\\
J_3a^\dagger =&sa^\dagger H\label{eq53c}\\
a J_3=&sHa \label{eq53d}\\
 \label{eq53e}[a ,a^\dagger ]=&(r-1)H+J_3
\end{eqnarray}
Using Eqs. (\ref{eq29a}) and (\ref{eq29b}) for this linear case we get: 
\begin{equation}\label{eq54}
\beta_{n+1}=s\alpha_n.
\end{equation}
\begin{equation}\label{eq55}
\alpha_{n+1}=r\alpha_n+\beta_n=r\alpha_n+s\alpha_{n-1}.
\end{equation}
The recurrence equation (\ref{eq55}) is a difference equation which generates a  Generalized Fibonacci Sequence. Hence, the sequence of eigenvalues of the $H$ operator follows a Generalized Fibonacci Series. We can write equations (\ref{eq54}) and (\ref{eq55}) as a two-dimensional linear map, which is more adequate 
in order to analyze the representation theory of the algebra. 
This map can be written as: 
\begin{eqnarray}
\label{eq56}
\left(
\begin{array}{c}
\alpha_{n+1} \\
\beta_{n+1}
\end{array}
\right) = 
\left(
\begin{array}{cc}
r & 1 \\
s & 0 
\end{array}
\right) 
\left(
\begin{array}{c}
\alpha_n \\
\beta_n
\end{array}
\right) .
\end{eqnarray}

\noindent

In \cite{cr}, the authors have analyzed, for one quantum number, the representation theory of the GHA using dynamical systems techniques . The representations were obtained analyzing the stability of the fixed-points and the cycles of the equation $\alpha_{n+1}=f(\alpha_n)$. In this work we have extended the previous effort and analyzed the stability of the fixed-points of the two dimensional system given by Eq. (\ref{eq56}). 

\subsection{Stability Analysis and Representation Theory}

The fixed-points of equation (\ref{eq56}) can be found by solving the system 
\begin{eqnarray}
\label{eq56-0}
\left(
\begin{array}{c}
\alpha^* \\
\beta^*
\end{array}
\right) = 
\left(
\begin{array}{cc}
r & 1 \\
s & 0 
\end{array}
\right) 
\left(
\begin{array}{c}
\alpha^* \\
\beta^*
\end{array}
\right) .
\end{eqnarray}
The solutions of the previous equation are  
$(\alpha^*,\beta^*)=(0,0)~\forall~ r,s  \in \mathbb{R}$ and 
$(\alpha^*,\beta^*)=(\alpha^*,s\alpha^*)$ for $r+s=1$, where
$r,s,\alpha^* \in \mathbb{R}$. The stability of these fixed points are given by 
the eigenvalues of the $2 \times 2$-matrix in Eq. (\ref{eq56-0}), which   
can be calculated by means of the 
characteristic equation, $\lambda^2-r\lambda-s=0$. These eigenvalues are 
independent of a specific point in the $(\alpha, \beta)$-space, as the recurrence 
given by Eq. (\ref{eq56}) is linear, and can be written as: 
\begin{equation}\label{eq56a}
\lambda_{\pm}=\frac{r\pm\sqrt{r^2+4s}}{2}.
\end{equation}

\noindent
The main possibilities are: \\
(i) the fixed point $(\alpha^*,\beta^*)$ is asymptotically stable if 
$|\lambda_{\pm}| < 1$; 
an initial  point $(\alpha_0,\beta_0)$, iterated using Eq. (\ref{eq56}), 
approaches  $(\alpha^*,\beta^*)$ as the 
number of iterations increases; \\ 
(ii) if at 
least one of the eigenvalues is, in modulus, bigger than one, the fixed point  
$(\alpha^*,\beta^*)$ 
is called unstable and the iterations of an initial point $(\alpha_0,\beta_0)$ 
by Eq. (\ref{eq56}) move this point far away the fixed point. \\
(iii) if both eigenvalues are, in modulus, equal to one - edge $BC$ 
in Figure (3) - 
we call  the fixed point $(\alpha^*,\beta^*) = (0,0)$ 
marginally stable . The iterations of an initial 
point in the $(\alpha, \beta)$-space will not move it far away the fixed point, 
but it will not approach the fixed point too as the iterations increase. 

In Figure (\ref{f1a}) we show the stability regions of the fixed points 
$(\alpha^*, \beta^*) = (0,0)$ and $(\alpha^*, \beta^*)  = (\alpha^*, s \alpha^*)$ 
in  the parameter space $(r,s)$. 

When $r + s \neq 1$ the fixed point $(\alpha^*,\beta^*)=(0,0)$  is the only 
fixed point in the $(\alpha,\beta)$ space. 
It is asymptotically stable for values of $r$ and $s$ inside the triangle $ABC$ 
and unstable for values of $r$ and $s$ outside this triangle. 
It is marginally stable on the edge $BC$ and, on the edge $AB$, 
where one eigenvalue ($\lambda_-$) is always equal to $-1$ and the other 
($\lambda_+$) is equal 
to $1+r$, with $r \in [-2,0]$, the iteration of any initial point $(\alpha_0,\beta_0)$ 
will approach a cycle two, with the line joining them crossing the fixed 
point $(0,0)$ in the $(\alpha,\beta)$-space. 

If $r + s = 1$, the dotted line in Figure (\ref{f1a}),  
there are a line of fixed points in the $(\alpha,\beta)$-space, 
$(\alpha^*,\beta^*)=(\alpha^*,(1-r) \alpha^*)$, with $\alpha^* \in \mathbb{R}$. 
All these fixed points are unstable if, in the line $r+s=1$, $r<0$ or 
$r>2$, i.e., for points in this line outside the edge $AC$ (see Figure (3)). 
For points $(r,s)$ in this line belonging to the edge $AC$, the 
associated eigenvalues 
are $\lambda_+ = +1$ and $\lambda_- = r-1$, with $r \in [0,2]$.  
The fixed points in the $(\alpha,\beta)$-space, 
$(\alpha^*, \beta^*)  = (\alpha^*, (1-r) \alpha^*)$, are 
stable in one direction, crossing the line of fixed points, 
associated with the eigenvalue 
$\lambda_- = r-1$, $r \in (0,2)$. 
The direction 
along the line of fixed points is marginally stable, 
associated with the eigenvalue $\lambda_+ = +1$. 

In the linear case, given the values of $r$ and $s$ (or, equivalently,  
 the values of $\lambda_+$ and $\lambda_-$, see Fig. 5),  
the possible values of $\alpha_0$ 
and $\beta_0$ for the representations of  the linear case of the algebra we are analyzing are presented in the appendix. 
If the two-dimensional map, Eq. (\ref{eq56}), has an unstable fixed point (i.e., if the values of 
$r$ and $s$ are outside the triangle of Fig. 3),  
the possible representations 
are infinite-dimensional and their corresponding spectra have no upper limit; 
the difference between two consecutive numbers in the 
corresponding sequence increases as the number of
iterations increases.
If this map has an asymptotically stable fixed point (i.e., if the values of $r$ 
and $s$ are inside the triangle of Fig. 3), the possible representations 
are also infinite-dimensional but they have 
superiorly-limited spectra; the 
difference between two consecutive numbers in the 
corresponding sequence decreases as the number of
iterations increases. 
The representations associated with marginally 
stable cases are more complex and, besides infinite-dimensional 
representations, it is also possible to  have finite-dimensional  ones. 
See the appendix for a discussion of the possible values of 
$\alpha_0$ and $\beta_0$ for these representations. 
 
\subsection{Fibonacci series}

If we assume $r=s=1$ the relation given by equation (\ref{eq55}) becomes
\begin{equation}\label{eq57}
\alpha_{n+1}= \alpha_n+\alpha_{n-1},
\end{equation}
which yields the usual Fibonacci sequence if we choose 
$(\alpha_0,\beta_0) = (1,0)$. 
The Fibonacci series have as  
eigenvalues $\lambda_{\pm} =  (1 \pm \sqrt{5})/2$. The $\lambda_{+} (>1)$ eigenvalue 
is the so-called golden-number. The single fixed-point $(\alpha^*, \beta^*) = (0,0)$ 
is, therefore, unstable. The representations are infinite dimensional and the 
sequences (depending on the initial values ($\alpha_0,\beta_0$) ) 
increase without an upper limit, as we know. 
Specific values of $\alpha_0$  and $\beta_0$ describing possible 
representations for this case will be discussed in the next section. 
The values  of $\alpha_n$ and $\beta_n$  for 
$r=s=1$ are shown in Table \ref{t1}. The eigenvalues $\alpha_n$ and $\beta_n$ 
provide the number of elements of kind ``A'' and ``B'', respectively, in the $n$-th 
inflation of the Fibonacci chain (Figure \ref{f1}).  Thus, the operator $a^\dagger $  acts 
as an ``inflation operator'' of the Fibonacci chain, while the $H$ eigenvalues  
provide the number of elements of the quasiperiodic chain generated at the 
$n$-th inflation.

\section{Fock space two-step Heisenberg algebra}

The Generalized Fibonacci Series (Eq. (\ref{eq55})) can be written by means of 
the so-called Binet formula: 
\begin{eqnarray}\label{eq59b}
\alpha_n=&\alpha_0\frac{ p^{n+1}- q^{n+1}}{ p - q}+ \beta_0\frac{ p^{n}- q^{n}}{ p - q}\\
\equiv &\alpha_0[n+1]_{p,q}+ \beta_0[n]_{p,q} ,
\end{eqnarray}
where $(p,q) \equiv (\lambda_+, \lambda_-)$ are the roots given by Eq.  (\ref{eq56a}), 
$n \geq 1$ and 
$[n]_{p,q}=\frac{p^n-q^n}{p -q}$ is the two-parameter $p,q$-number 
(Gauss Number) \cite{cj}. As $H$ is the Hamiltonian, the choice of $\beta_0$ 
is not completely free inasmuch as we should guarantee that $\alpha_n \geq \alpha_0$ 
for any $n > 0$ (the energy of the ground state must be the lowest one). This 
condition implies that, once the value of $\alpha_0$ had been chosen, 
the allowed values of $\beta_0$ should satisfy the relation
\begin{equation}
\label{betaoa}
\beta_0  [n]_{p,q} \geq ( 1-[n+1]_{p,q} ) \, \alpha_0 ,
\end{equation}
for any value of $n$, $n>0$. 
In particular, if $ [n]_{p,q} > 0$ for any value of $n>0$, 
the condition to be satisfied by $\beta_0$ can be written as:

\begin{equation}
\label{betao}
\beta_0 \geq \frac{ 1-[n+1]_{p,q} }{ [n]_{p,q}} \, \alpha_0 , 
\end{equation}
for any value of $n$, $n>0$. 

The solution for $\beta_0$ for the whole parameter space will be described 
in the appendix. For the specific case of Fibonacci chain ($r=s=1$) it is simple 
to see that the term $(1-[n+1]_{p,q} ) /  [n]_{p,q}$ varies from 
$-(1+ \sqrt{5})/2$ up to $0$. Then, $\alpha_0 \geq 0$ implies 
$\beta_0 \geq 0$ and $\alpha_0 < 0 $ implies that  
$\beta_0 \geq ((1+ \sqrt{5}\,)/2) | \, \alpha_0|$.

We can show that there exist another Casimir Operator $C^{(2)}$ when  
$f(x)$ and $g(x)$ are linear functions:
\begin{equation}\label{eq59c}
C^{(2)}=a a^\dagger -\alpha_0([N+2]_{p,q}-1)- \beta_0([N+1]_{p,q}-1),
\end{equation}
where $[N]_{p,q}=\frac{ p^N- q^{N}}{ p - q}$ is the $p,q-$deformed number operator and $N$ is the usual number operator, defined in Fock space as 
\begin{equation}
\label{n}
N|n\rangle=n|n\rangle \, .
\end{equation}

Using Eq.  (\ref{eq59b}), it is possible to reduce the algebraic structure 
(\ref{eq53a}-\ref{eq53e}) to a Heisenberg-like structure composed by the 
usual operators $A$, $A^\dagger$ and $N$. From Eq. (\ref{eq59b}) $H$ can 
be written in Fock space as:  
\begin{equation}
\label{eq60} 
H\rightarrow \alpha_0\frac{p^{N+1}-q^{N+1}}{p-q}+ \beta_0\frac{p^{N}-q^{N}}{p-q} \, ,
\end{equation}
and, using Eq. (\ref{eq54}), $J_3$ can be written as  
\begin{equation}
\label{eq61}
J_3\rightarrow s \left( \alpha_0\frac{p^{N}-q^{N}}{p-q}+\beta_0\frac{p^{N-1}-q^{N-1}}{p-q}
\right).
\end{equation}
 Note that the effect of the application of $H$ and $J_3$ on $|n\rangle$ is exactly the 
 same as if we use Eqs. (\ref{eq60}) and (\ref{eq61}) on  $|n\rangle$.
Substituting $H$ and $J_3$ in Eq. ({\ref{eq53e}}) by Eqs.  
(\ref{eq60}) and (\ref{eq61}) and using Eqs. (\ref{eq28a}), (\ref{eq28b}) and 
(\ref{n}) we get:  
\begin{eqnarray}\label{eq62a}
[N,a^\dagger]&=&a^\dagger,\\\label{eq62b}
[N,a]&=&-a,\\\label{eq62c}
[a,a^\dagger]&=&(r-1)\left( \alpha_0[N+1]_{p,q}+ \beta_0[N]_{p,q} \right) +
\nonumber\\
&+&  s \left( \alpha_0[N]_{p,q}+ \beta_0[N-1]_{p,q} \right).
\end{eqnarray}
The above algebraic structure is the Fock space form of a two-step Heisenberg algebra. 
Notice that the algebraic structure described in Eqs. (\ref{eq53a}-\ref{eq53e}) correspond 
to Eqs. (\ref{eq62a}-\ref{eq62c}) plus the Hamiltonian 
$H =\alpha_0[N+1]_{p,q}+ \beta_0[N]_{p,q}$, whose 
eigenvalues follow the generalized Fibonacci sequence.
  
Remembering that $p . q=-s$ and $p+q=r$ ($p$ and $q$ are roots of $x^2-rx-s=0$), the  right hand side of Eq. (\ref{eq62c}) can be written as
\begin{equation}\label{eq63}
(p+q-1)(\alpha_0[N+1]_{p,q} + \beta_0[N]_{p,q})-pq(\alpha_0[N]_{p,q} +
 \beta_0[N-1]_{p,q}),
\end{equation}
which, after some algebra allows us to write Eq. (\ref{eq62c}) as:
\begin{eqnarray}\label{eq63a}
[a,a^\dagger]&=\alpha_0\left( [N+2]_{p,q}-[N+1]_{p,q}\right) + 
\beta_0 \left( [N+1]_{p,q} - [N]_{p,q} \right).
\end{eqnarray}
Considering $\alpha_0=1$ and $\beta_0=0$ in Eq. (\ref{eq63a}), Eqs. (\ref{eq62a}-\ref{eq62b}) 
and (\ref{eq63a}) can be related to  
the $(p,q)$-oscillator, reference \cite{cj},  which was used in nuclear spectroscopy \cite{bmk}. 
Thus, the $H$ eigenvalues of the $(p,q)$-oscillator can be written as a particular case of the generalized Fibonacci numbers with ``initial conditions" $\alpha_0=1$ and $\beta_0=0$. In our ``extended algebra", the $H$ eigenvalues are written as Generalized Fibonacci numbers where the initial values $\alpha_0$ and $\beta_0$ can assume any value since that $\alpha_0 \leq \alpha_n$ for any $n>0$. Thereby, each set of initial values 
($\alpha_0,\beta_0$) define a new representation.

\section{Final comments}

 We have constructed a Heisenberg-type algebra having as generators 
 the Hamiltonian, the step-operators and an auxiliary operator. 
 Moreover, this structure depends on two analytic 
 functions $f(x)$ and $g(x)$. Since the energy spectra of the systems 
 described by this algebra depend on the two previous energy levels, 
 this algebra was called generalized two-step Heisenberg algebra. 
  
 This algebra describe quantum systems having eigenvalues of the 
 operator $H$ (Hamiltonian) analytically depending on the two previous 
 eigenvalues,  
 i. e.,  
 $\epsilon_{i+1} = h(\epsilon_{i}, \epsilon_{i-1})$, where 
 $h: \Re^2 \rightarrow \Re$ 
 is a function 
 and $\epsilon_{i}$ is the eigenvalue of 
 $H$ acting on the eigenstate $i$, i. e., $H | i \rangle = \epsilon_{i} | i \rangle$. 
 A quantum system having the Hamiltonian eigenvalues obeying the Fibonacci 
 sequence (quasi-periodic systems) could, for example, be described by this algebra. 
 Possible spectra generated by this algebra can be seen in Figure \ref{level}, 
 showing that the typical possible spectra found in nature can be generated 
 by this algebraic structure.
 
 We have discussed the representations of this algebra and we have classified 
 in detail the possible representations for linear $f(x)$ and $g(x)$. 
 To perform this classification we 
 have used tools from dynamical system techniques as 
the analysis of the stability of the fixed 
 points of the characteristic functions of the algebra, $f(x)$ and $g(x)$. 
  
In \cite{jnc}, an algebraic phenomenological approach, based on the GHA, 
was implemented for the $C O_2$ molecule. We think that the algebra 
constructed here could be used to generalize the mentioned approach 
to more general spectra.  

 We also consider mathematically interesting to analyse in detail the 
 representations  of this algebra  at least  for simple nonlinear examples of 
 the characteristic  functions. 
 
 \section*{Appendix} 

In this appendix we discuss the representations associated with the linear case.   
In order to classify the possible values of $\alpha_0$ and $\beta_0$ 
according to the value of the eigenvalues 
$\lambda_+$ (or $p$) and $\lambda_-$ (or $q$) we divide the plane 
($\lambda_-, \lambda_+$) in regions labeled by the numbers $I$, $II$, $III$, $IV$, $V$ and 
$VI$, see figure \ref{lambda}.  As, by definition, $\lambda_+ \ge \lambda_-$, only the semi-plane 
equal and above the diagonal line at $45^o$ in this plane makes sense. The representations 
of the extended algebra can be obtained providing $\alpha_0$ and $\beta_0$ satisfy 
the conditions given by Eqs. (\ref{betaoa}) and (\ref{betao}). Analysing these equations 
for the regions labeled from $I$ to $VII$ in Figure \ref{lambda}, it is possible to obtain 
representations for the extended algebra since $\alpha_0$ and $\beta_0$ satisfy 
the following conditions (the results for regions IV, VI and VII are numerical): 

\vspace{0.3cm}

\noindent
Region I: $\lambda_+ >1$ and $ -1 \le \lambda_- \le  \lambda_+$\\
a) if $\alpha_0 \ge 0$ $\rightarrow$ $\beta_0 \ge (1 - \lambda_+ -  \lambda_-) \alpha_0$ \\
b) if $\alpha_0 < 0$ $\rightarrow$ $\beta_0 \ge  \lambda_+ | \alpha_0 |$
\vspace{0.2cm}

\noindent
Region II: $0 < \lambda_+ <1$ and $ - \lambda_+ < \lambda_- <    \lambda_+$ \\
$\alpha_0 \le 0$ $\rightarrow$ $\beta_0 \ge (\lambda_+ + \lambda_- -1) |\alpha_0|$

\vspace{0.2cm}

\noindent
Region III: $-1 < \lambda_+ < 1$ and $-1 < \lambda_- < \mbox{min} (-\lambda_+,\lambda_+)$\\
$\alpha_0 = 0$  and $\beta_0 \ge 0$

\vspace{0.2cm}

\noindent
Region IV: $ \lambda_+ > 1$ and $ -\lambda_+ < \lambda_- < -1$\\
a) if $\alpha_0 \ge 0$ $\rightarrow$ $\beta_0 \ge (1 - \lambda_+ +  |\lambda_-|) \alpha_0$ \\
b) if $\alpha_0 < 0$ $\rightarrow$ $\beta_0 \ge 
(\frac{\lambda_+^2 + |\lambda_-|^2 - \lambda_+ |\lambda_-| -1}{\lambda_+ - |\lambda_-|}) 
|\alpha_0|$

\vspace{0.2cm}

\noindent
Region V: $0 < \lambda_+ < -\lambda_-$ and $\lambda_- < \mbox{min}(-\lambda_+,-1)$\\
In this region there is no simple general analytical solution for the values 
of $\alpha_0$ and $\beta_0$ and the possible representations should be 
computed case by case.

\vspace{0.2cm}

\noindent
Region VI: $\lambda_- < \lambda_+ < 0$ and $\lambda_- < \mbox{min}(-1,\lambda_+)$ \\
a) if $\alpha_0 \ge 0$ $\rightarrow$ $\beta_0 \ge (1 + |\lambda_+| +  |\lambda_-|) \alpha_0$ \\
b) if $\alpha_0 < 0$ $\rightarrow$ $\beta_0 \ge
(\frac{\lambda_+^2 + \lambda_-^2 - |\lambda_+| |\lambda_-| -1}
{|\lambda_-| - |\lambda_+|}) |\alpha_0|$ 

\vspace{0.2cm}

\noindent
Region VII: $\lambda_+ < -1$, $\lambda_- < -1$ and $|\lambda_-| > |\lambda_+|$. \\
a) if $\alpha_0 \ge 0$ $\rightarrow$ $\beta_0 \ge (1 + |\lambda_+| +  |\lambda_-|) \alpha_0$ \\
b) if $\alpha_0 < 0$ $\rightarrow$ $\beta_0 \ge - |\lambda_-| | \alpha_0 |$ 

\vspace{0.2cm}

\noindent
The possible expressions for $\alpha_0$ and $\beta_0$ belonging to the 
boundaries between these regions are the following: 

\vspace{0.3cm}

\noindent
i) Boundary between region $I$ and $II$: \\
$\beta_0 \ge - \lambda_- \, \alpha_0$  for $\alpha_0 \in \Re$

\vspace{0.3cm}

\noindent
ii) Boundary between region $I$ and $IV$: \\
The same expressions for $\alpha_0$ and $\beta_0$ allowed for region $I$ 
with $\lambda_- = -1$.
 
 \vspace{0.3cm}

\noindent
iii) Boundary between region $II$ and $III$: \\
The same expressions for $\alpha_0$ and $\beta_0$ allowed for region $III$ 
with $\lambda_- = - \lambda_+$.

 \vspace{0.3cm}

\noindent
iv) Boundary between region $IV$ and $V$: \\
$\beta_0 \ge \alpha_0$.

 \vspace{0.3cm}

\noindent
v) Boundary between region $V$ and $VI$: \\
The same expressions for $\alpha_0$ and $\beta_0$ allowed for region $VI$ 
with $\lambda_+ = 0$.

 \vspace{0.3cm}

\noindent
vi) Boundary between region $III$ and $V$: \\
a) If $\alpha_0 \ge 0$ $\rightarrow \beta_0 \ge (2 + \lambda_+) \, \alpha_0$ \\
b) If $\alpha_0 < 0 \rightarrow \beta_0 \ge \lambda_+ \, |\alpha_0|$.
 
 \vspace{0.3cm}

\noindent
vii) Boundary between region $III$ and $VI$: \\
 The same expressions for $\alpha_0$ and $\beta_0$ allowed for region $VI$ 
with $\lambda_- = -1$.
 
  \vspace{0.3cm}

\noindent
viii) Boundary between region $VI$ and $VII$: \\
 The same expressions for $\alpha_0$ and $\beta_0$ allowed for region $VI$ 
with $\lambda_+ = -1$.

\begin{table}
\large
\centering
\begin{tabular}{|c|c|c|}\hline
$n$&$\beta_n=\alpha_{n-1}$    &  $\alpha_n=\alpha_{n-1}+\beta_{n-1}$  \\ \hline\hline
0 & $\beta_0$            & $\alpha_0$  \\\hline
1 & $\alpha_0$           & $\alpha_0+\beta_0$\\\hline
2 & $\alpha_0+\beta_0$  & $2\alpha_0+\beta_0$\\\hline
3 & $2\alpha_0+\beta_0$ & $3\alpha_0+2\beta_0$  \\ \hline
4 & $3\alpha_0+2\beta_0$ & $5\alpha_0+3\beta_0$  \\ \hline
\end{tabular}
\caption{Eigenvalues of $H$ and $J_3$ operators when $r=s=1$.}
\label{t1}
\end{table}

\begin{center}
\begin{figure}
\setlength{\unitlength}{1.0 cm}
\begin{picture}(10,.8)
        \put(1,0.6){\line(1,0){10}}
        \put(1,0.1){\small$A\rightarrow AB \rightarrow ABA \rightarrow ABAABA \rightarrow ABAABAABA\rightarrow \ldots$}
        \put(1,0){\line(1,0){10}}
\end{picture}
\caption{The  Fibonacci Chain. The substitution rule for build this chain is: A$\rightarrow$AB and B$\rightarrow$A.}
\label{f1}
\end{figure}
\end{center} 

\begin{figure}
\setlength{\unitlength}{1 cm}
\begin{picture}(10,.8)
        \put(1,0.6){\line(1,0){11}}
        \put(1,0.1){\small$A\rightarrow ABA \rightarrow ABAAABA \rightarrow ABAAABAABAABAAABA\rightarrow \ldots$}
        \put(1,0){\line(1,0){11}}
\end{picture}
\caption{A Fibonacci-like chain. The substitution rule for build this chain is: A$\rightarrow$ABA and B$\rightarrow$A.}
\label{f2}
\end{figure}

 \begin{figure}
 \begin{center}
 \includegraphics[scale=0.8,angle=0]{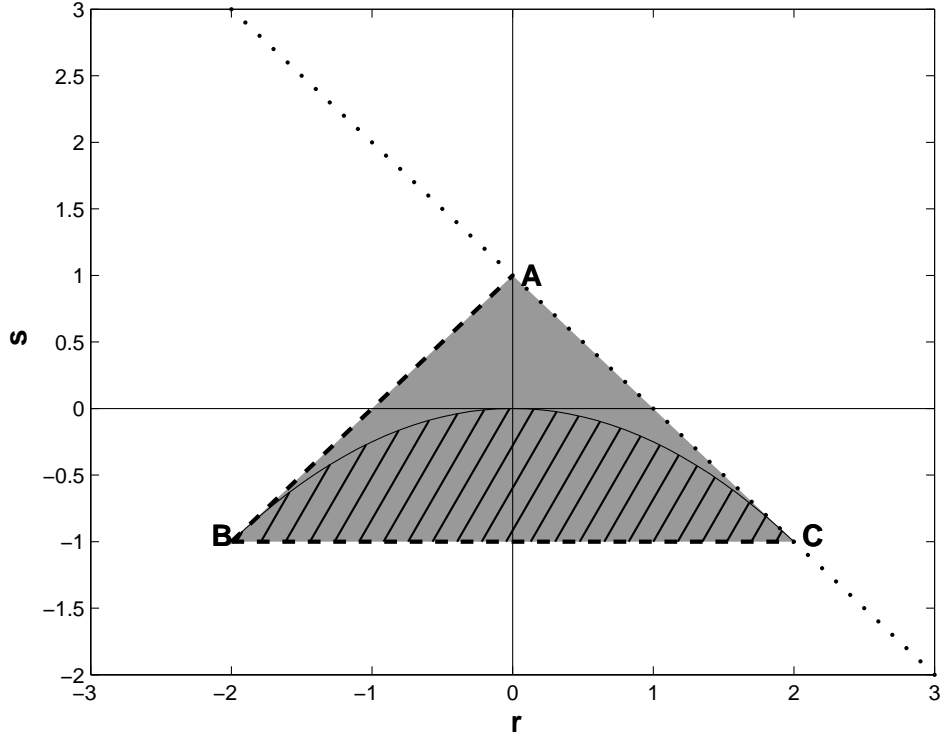}
 \end{center}
 \caption{Stability regions in the parameter $(r,s)$-space. The asymptotically stable region 
 of the $(\alpha=0,\beta = 0)$ 
 fixed-point, i.e., $|\lambda_{\pm}| < 1$,  lies inside the triangle $ABC$ 
(edges are not included). Outside the triangle $ABC$ the  $(0,0)$ fixed-point in the 
$(\alpha,\beta)$-space 
is unstable (at least one eigenvalue has modulus greater than one). The hached area corresponds to the asymptotically stable region where the eigenvalues are complex. The marginally stable region 
of the $(0,0)$ fixed-point lies on the dashed edges. 
  On the edge $\mathrm{B} \mathrm{C}$ both eigenvalues $\lambda_{\pm}$ are complex, 
  with unitary modulus.  
  The dotted line is the region in the parameter space allowing the existence of the 
  fixed-points $(\alpha^*,s\alpha^*)$ in the dynamical space $(\alpha,\beta)$. 
  These fixed points are marginally stable 
 for values of the parameters $(r,s)$ on the edge $AC$ and unstable outside.
 The eigenvalues at the points $A$, $B$, and $C$ are, respectively, $(1,-1)$, $(-1,-1)$ and 
 $(1,1)$.}
 \label{f1a}
\end{figure}

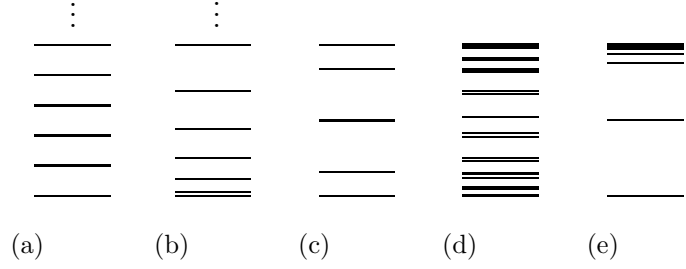
\begin{figure}

    \centering

    \subfigure[\label{a}]{
\setlength{\unitlength}{0.05 cm}
\begin{picture}(30,40)
        \put(10,45){\vdots}
        \put(1,40){\line(1,0){20}}
        \put(1,32){\line(1,0){20}}
        \put(1,24){\line(1,0){20}}
        \put(1,16){\line(1,0){20}}
        \put(1,8){\line(1,0){20}}
        \put(1,0){\line(1,0){20}}
\end{picture}}
 \subfigure[\label{b}]{
\setlength{\unitlength}{0.005 cm}
\begin{picture}(300,400)
        \put(100,450){\vdots}
        \put(1,400){\line(1,0){200}}
        \put(1,278){\line(1,0){200}}
        \put(1,178){\line(1,0){200}}
        \put(1,100){\line(1,0){200}}
        \put(1,44){\line(1,0){200}}
        \put(1,10){\line(1,0){200}}
        \put(1,0){\line(1,0){200}}
\end{picture}}
 \subfigure[\label{c}]{
\setlength{\unitlength}{0.005 cm}
\begin{picture}(300,400)
        \put(1,400){\line(1,0){200}}
        \put(1,336){\line(1,0){200}}
        \put(1,200){\line(1,0){200}}
        \put(1,64){\line(1,0){200}}
        \put(1,0){\line(1,0){200}}
\end{picture}}
\subfigure[\label{d}]{ 
\setlength{\unitlength}{0.0005 cm}
\begin{picture}(3000,4000)
        \put(1,4000){\line(1,0){2000}}
        \put(1,3934){\line(1,0){2000}}
        \put(1,3909){\line(1,0){2000}}
        \put(1,3639){\line(1,0){2000}}
        \put(1,3588){\line(1,0){2000}}
        \put(1,3352){\line(1,0){2000}}
        \put(1,3286){\line(1,0){2000}}
        \put(1,2764){\line(1,0){2000}}
        \put(1,2683){\line(1,0){2000}}
        \put(1,2084){\line(1,0){2000}}  
        \put(1,1649){\line(1,0){2000}}
        \put(1,1563){\line(1,0){2000}}
        \put(1,995){\line(1,0){2000}}
        \put(1,920){\line(1,0){2000}}
        \put(1,580){\line(1,0){2000}}
        \put(1,461){\line(1,0){2000}}
        \put(1,219){\line(1,0){2000}}
        \put(1,180){\line(1,0){2000}}
        \put(1,10){\line(1,0){2000}}
        \put(1,0){\line(1,0){2000}}
        \end{picture}}
\subfigure[\label{e}]{
\setlength{\unitlength}{0.0005 cm}
\begin{picture}(3000,4000)
        \put(1,4000){\line(1,0){2000}}  
        \put(1,3998){\line(1,0){2000}}
        \put(1,3996){\line(1,0){2000}}
        \put(1,3992){\line(1,0){2000}}
        \put(1,3984){\line(1,0){2000}}
        \put(1,3968){\line(1,0){2000}}
        \put(1,3937){\line(1,0){2000}}
        \put(1,3875){\line(1,0){2000}}
        \put(1,3750){\line(1,0){2000}}
        \put(1,3500){\line(1,0){2000}}
        \put(1,2000){\line(1,0){2000}}
        \put(1,0){\line(1,0){2000}}
        \end{picture}}

\caption{Examples of energy levels for various parameters of the algebra. 
In $(a)$, for $r=2$ and 
$s=-1$, the levels are evenly spaced. In $(b)$,  for $r=3$ and $s=-2$ the level spacing 
is increasing. In $(c)$, the levels are periodic for  $r=2 \cos \tau $ and $s=-1$, where 
$\tau = 2 \pi /k$, with $k = 3,4, \cdots$. 
In $(d)$, the level spacing form a dense set for $r=2 \cos \tau $ and $s=-1$, where 
$\tau = 2 \pi \gamma$, with $\gamma$ being an irrational number. In $(e)$, 
$r=3/2$ and $s=-1/2$, the level spacing is decreasing.
}\label{level}
\end{figure}

\begin{figure}
\begin{center}
\includegraphics[width=3in]{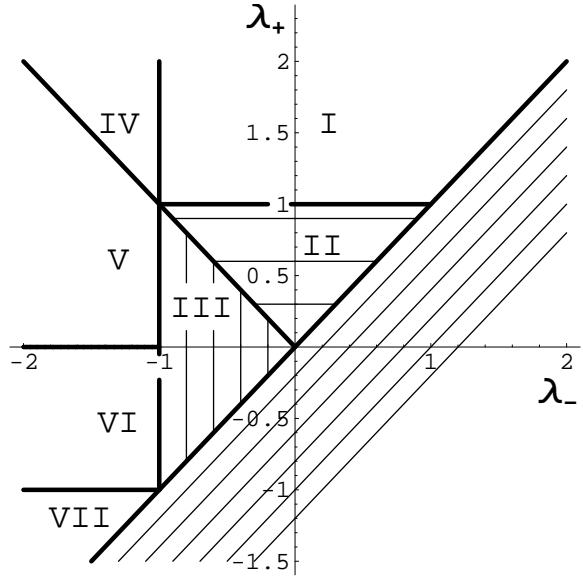}
\caption{Regions in the $(\lambda_-, \lambda_+)$-plane associated with different   
conditions on $(\alpha_0,\beta_0)$ allowing lowest vacuum eigenstate representations. }
\label{lambda}
\end{center}
\end{figure}

\end{document}